\documentclass[aps,pra,floatfix,showpacs,twocolumn]{revtex4}
\usepackage{graphicx} \usepackage{amsmath} \usepackage{amssymb}
\usepackage{color} \usepackage[hyperindex,breaklinks]{hyperref} \usepackage{epstopdf}
\DeclareMathOperator{\tr}{tr}
\DeclareMathOperator{\diag}{diag}
\newcommand{\comment}[1]{}
\newcommand{\BEQ}{\begin{equation}}
\newcommand{\EEQ}{\end{equation}}
\newcommand{\BEA}{\begin{eqnarray}}
\newcommand{\EEA}{\end{eqnarray}}
\renewcommand{\d}{{\rm d}}
\newcommand{\eps}{\varepsilon}

\renewcommand{\P}{{\cal P}}

\begin{document}
\draft
\title{Carnot cycle at finite power: Attainability of maximal efficiency}

\author{Armen\ E.\ Allahverdyan,$^{1}$ Karen\ V.\ Hovhannisyan,$^{2,\,1}$ Alexey\ V.\ Melkikh,$^{3}$ and Sasun\ G.\ Gevorkian$^{4}$}
\affiliation{$^{1}$Yerevan Physics
  Institute, Alikhanian Brothers Street 2, Yerevan 375036, Armenia}

\affiliation{$^{2}$ICFO-Institut de
  Ci\`encies Fot\`oniques, Mediterranean Technology Park, 08860
  Castelldefels (Barcelona), Spain}

\affiliation{$^{3}$Ural Federal University,
  Mira Street 19, Yekaterinburg 620002, Russia}

\affiliation{$^{4}$Institute of Physics,
  Academia Sinica, Nankang, Taipei 11529, Taiwan}

\begin{abstract} We want to understand whether and to which extent the
  maximal (Carnot) efficiency for heat engines can be reached at a
  finite power. To this end we generalize the Carnot cycle so that it
  is not restricted to slow processes. We show that for realistic
  (i.e. not purposefully-designed) engine-bath interactions, the
  work-optimal engine performing the generalized cycle close to the
  maximal efficiency has a long cycle time and hence vanishing
  power. This aspect is shown to relate to the theory of computational
  complexity. A physical manifestation of the same effect is the
  Levinthal's paradox in the protein folding problem. The resolution
  of this paradox for realistic proteins allows to construct engines
  that can extract at a finite power 40 \% of the maximally possible
  work reaching 90 \% of the maximal efficiency. For
  purposefully designed engine-bath interactions, the Carnot
  efficiency is achievable at a large power.
\end{abstract}



\maketitle

Reciprocating heat engines extract work operating cyclically between two
thermal baths at temperatures $T_1$ and $T_2$ ($T_1>T_2$) \cite{callen}.
They have two basic characteristics: {\it (i)} efficiency, $\eta=W/Q_1$,
is the work $W$ extracted per cycle divided by the heat input $Q_1$ from
the high-temperature bath. {\it (ii)} Power $W/\tau$, where $\tau$ is
the cycle duration. Both these quantities have to be large for a good
engine: if $\eta$ is small, lot of energy is wasted; if the power is
small, no sizable work is delivered over a reasonable time
\cite{callen}.

The second law establishes the Carnot efficiency $\eta_{\rm
C}=1-\frac{T_2}{T_1}$ as an upper bound for $\eta$ \cite{callen}. The
Carnot cycle reaches the bounding value $\eta_{\rm C}$ in the (useless)
limit, where the power goes to zero \cite{callen}. Conversely, realistic
engines are not efficient, since they have to be powerful, e.g. the
efficiency of Diesel engines amounts to 35--40 \% of the maximal value.
This {\it power-efficiency dilemma} motivated a search for the
efficiency that would generally characterize the maximal power regime.
One candidate for this is the Curzon-Ahlborn efficiency $\eta_{\rm CA}
=1-\sqrt{{T_2}/{T_1}}$ \cite{CA}, which is however crucially tied to the
linear regime $T_1\approx T_2$ \cite{CA_van,CA_vanvan}. Beyond this
regime $\eta_{\rm CA}$ is a lower bound of $\eta$ for a class of model
engines \cite{engine}. Several recent models for the efficiency at the
maximal power overcome $\eta_{\rm CA}$ with $\eta^*=\frac{\eta_{\rm
C}}{2-\eta_{\rm C}}$ \cite{tarm1}.

As argued in \cite{engine,tarm2,tarm3}, the maximal power regime
allows for the Carnot efficiency, at least for certain models. But it
is currently an open question whether the maximal efficiency is
attained under realistic conditions (see e.g. \cite{tarm4} versus
\cite{tarm2}), and how to characterize the very {\it realism} of those
conditions. Even more generally: what is the origin of the
power-efficiency dilemma? We answer these questions by analyzing a
generalized Carnot cycle, which in contrast to the original Carnot
cycle is not restricted to slow processes. We now summarize our
answers.

{\it (1)} When the $N$-particle engine operates at the maximal work
extracted per cycle, its efficiency reaches the Carnot bound
$\eta_{\rm C}$ for $N\gg 1$, while the cycle time is given by the
relaxation time of the engine. The maximal work and the Carnot
efficiency are achieved due to the flat energy landscape of the
engine. For realistic engine-bath interactions this energy landscape
leads to a very long [${\cal O}(e^N)$] relaxation time nullifying the
power. By realistic we mean interactions that are independent from the
engine Hamiltonian. If we assume a proper tuning between engine-bath
interaction and the engine Hamiltonian, the relaxation time scales as
${\cal O}(\sqrt{N})$, and the maximal efficiency is achievable in the
limit $N\gg 1$ at a large power ${\cal O}(\sqrt{N})$.

{\it (2)} The relaxation of the optimal engine under realistic
interactions relates to an important problem of searching an
unstructured database for a marked item, where each energy level refers
to a database item. This task is computationally complex, i.e. even the
most powerful quantum algorithms resolve it in ${\cal O}(e^{N/2})$
time-steps \cite{farhi}. Hence the power-efficiency dilemma relates to
computational complexity. The same effect can be reformulated as the
Levinthal's paradox of the protein folding problem: if the majority of
unfolded states of a protein are assumed to have the same (free) energy,
the folding time is very long \cite{zwan}.

{\it (3)} A scenario of resolving the Levinthal's paradox proposed in
protein science shows the way of constructing sub-optimal engines that
operate at a reasonably large values of work, power and
efficiency. These sub-optimal engines function as model proteins, but
they are restricted to mesoscopic scales $N\sim 100$; otherwise the
relaxation time is again large.  Sacrificing some 50--60\% of the
maximal possible work leads to a reasonable cycle times with the
efficiency that achieves some 90 \% of the maximal (Carnot)
efficiency.

\textit{\textbf{Carnot cycle and its generalization.}} Recall that the
Carnot cycle consists of four slow, consecutive pieces \cite{callen}:
thermally-isolated -- isothermal -- thermally-isolated -- isothermal.
Four times slow brings in the vanishing power stressed above; see
additionally section I of the Supplementary Material. Since the overall process is a
quasi-equilibrium one, the external fields that extract work from the
engine act on it during all the four pieces. One deduces for the
isothermal parts: $Q_1=T_1\Delta S$ and $Q_2=T_2\Delta S$, where $Q_1$
($Q_2$) is the heat taken from (put into) the $T_1$-bath ($T_2$-bath),
and $\Delta S>0$ is the entropy change. Since the work extracted is
$W=Q_1-Q_2$, $\eta$ equals to its maximal value $\eta_{\rm
C}=1-\frac{T_2}{T_1}$ \cite{callen}.

We keep the two isothermal and two thermally isolated pieces of the
original Carnot cycle, but do not force them to be slow. In addition,
the external fields will act only during the thermally isolated
stages. Isothermal pieces amount to free relaxation. Due to these
points, we can analyze the engine functioning from the energy
conservation. We study the engine via quantum mechanics on a finite
Hilbert space, because this reduces the problem to a combinatorial
optimization. The final results are interpreted classically and can be
also obtained by discretizing the Hamiltonian classical dynamics over
phase-space cells.

{\bf 0.} The engine E with the Hamiltonian $H_1$ starts
in an equilibrium state at temperature $T_1$
described by the density matrix
\BEA
\label{0}
\rho(0)=\rho_1 = e^{-\beta_1H_1}/({\rm tr}\, e^{-\beta_1H_1}), ~~ \beta_1=1/T_1.
\EEA

{\bf 1.} Between times $0$ and $\tau$, E undergoes a thermally isolated
process with a time-dependent Hamiltonian $H_{12}(t)$ and the unitary
evolution $\rho(\tau)=U_{1 2}\rho(0)U_{1 2}^\dagger$:
\BEA
H_{1 2}(0)=H_1, ~~ H_{1 2}(\tau)=H_2, ~~
U_{1 2}={\cal T}e^{-i\int_0^\tau \d s H_{1 2}(s)},
\EEA
where ${\cal T}$ means chronological ordering.
The work taken out of E is determined by
energy conservation (see \cite{callen} and
section II of the Supplementary Material)
\BEA
\label{w1}
W_{1}= {\rm tr}[H_1\rho_1-H_2 U_{1 2}\rho_1 U_{1 2}^\dagger ].
\EEA

{\bf 2.} Then E is attached to the $T_2$-bath and after relaxation time
$\tau_{\rm r}$ its density matrix becomes
\BEA
\rho(\tau+\tau_{\rm r})=\rho_2= e^{-\beta_2H_2}/({\rm tr}\, e^{-\beta_2H_2}).
\label{00}
\EEA
The heat that came to E from the $T_2$-bath is
\BEA
Q_{2}= {\rm tr}[H_2\rho_2- H_2 U_{12}\rho_1 U_{12}^\dagger].
\EEA

{\bf 3.} E undergoes another thermally isolated process
\BEA
H_{21}(0)=H_2, ~~ H_{21}(\tau)=H_1, ~~
U_{21}={\cal T}e^{-i\int_0^\tau \d s H_{21}(s)},
\EEA
completing the cycle with respect to the Hamiltonian.
The work taken out of E reads
\BEA
\label{w2}
W_{2}= {\rm tr}[H_2\rho_2 - H_1 U_{21}\rho_2 U_{21}^\dagger ].
\EEA

{\bf 4.} Finally, E is attached to the $T_1$-bath ($T_1>T_2$) and relaxes to
$\rho_1$ thereby completing the cycle; see (\ref{0}). The heat that
came to E from the $T_1$-bath is
\BEA
\label{q1}
Q_{1}= {\rm tr}[H_1\rho_1- H_1 U_{21}\rho_2 U_{21}^\dagger].
\EEA
To stress the differences with the original Carnot cycle: {\it (i)} the cycle
time $2(\tau+\tau_{\rm r})$ need not be much larger than the relaxation
time $\tau_{\rm r}$. {\it (ii)} The cycle is out of equilibrium. {\it
(iii)} The work source and the bath never act simultaneously; either one
acts or another. Hence heat and work are deduced from the
energy conservation.

We did not count the work necessary for switching the system-bath
interaction on and o ff, because we assume that it does not contribute
to the total work budget (e.g. since it is weak).

\textit{\textbf{Maximization of work.}} We maximize the full extracted
work $W=W_1+W_2$ over $H_1$, $H_2$, $U_{12}$, $U_{21}$ for fixed
$T_1>T_2$ and a fixed number $n+1$ of energy levels of E. The lowest
energies of $H_1$ and $H_2$ can be set to zero. Introduce the
eigen-resolution of $H_\alpha$
\BEA
\label{min}
H_\alpha={\sum}_{k=2}^{n+1} \epsilon_k^{[\alpha]}|k^{[\alpha]}
\rangle\langle k^{[\alpha]}|,~~ \alpha=1,2.
\EEA
The full work $W=W_1+W_2$ reads from (\ref{w1}, \ref{w2})
\BEA
\label{bela}
W&=& {\sum}_{\alpha=1}^2{\sum}_{k=2}^{n+1} p_k^{[\alpha]}\epsilon_k^{[\alpha]} \\
&-& {\sum}_{k,l=2}^{n+1} \left[ p_k^{[2]}\epsilon_l^{[1]} C^{[21]}_{kl}
+ p_k^{[1]}\epsilon_l^{[2]} C^{[12]}_{kl}\right], \label{ballast}
\EEA
where $\{p_k^{[\alpha]}\}_{k=1}^{n+1}$ are eigenvalues of
$\rho_{\alpha}$ given by (\ref{0}, \ref{00}, \ref{min}), and where
\BEA
C^{[\alpha\gamma]}_{kl}\equiv |\langle k^{[\alpha]}|U_{\alpha\gamma}|
l^{[\gamma]}\rangle|^2, ~~~
(\alpha,\gamma)=(1,2), \, (2,1).
\label{buka}
\EEA
$C^{[\alpha\gamma]}_{kl}$ are doubly stochastic matrices:
$\sum_{k=1}^{n+1} C^{[\alpha\gamma]}_{kl}=\sum_{l=1}^{n+1}
C^{[\alpha\gamma]}_{kl}=1$. Such a matrix $C_{kl}$
can be represented as a convex sum of permutation matrices (Birkhoff's
theorem) \cite{olkin}:
$C_{kl}=\sum_{\delta}\lambda_\delta\Pi^{[\delta]}_{kl}$, where
$\lambda_\delta\geq 0$, $\sum_{\delta}\lambda_\delta=1$, and where
$\Pi^{[\delta]}$ permutes the components of any vector on which it
acts. Hence we can maximize $W$ in (\ref{bela}, \ref{ballast}) over
$\{\lambda_\delta\}$. The optimal $C^{[12]}_{kl}$ and $C^{[21]}_{kl}$
amount to permutation matrices, since $\{\lambda_\delta\}$ enter
linearly into $W$. Without loss of generality we can assume
$\epsilon_1^{[\alpha]}\leq...\leq \epsilon_{n+1}^{[\alpha]}$ and hence
$p_1^{[\alpha]}\geq...\geq p_{n+1}^{[\alpha]}$. Then the optimal permutations
$C^{[12]}_{kl}$ and $C^{[21]}_{kl}$ are unit matrices; see
(\ref{ballast}). In contrast to
the original Carnot cycle, the optimal thermally isolated processes
can be realized as sudden (quick) changes of the Hamiltonian
eigenvalues without changing the eigenvectors. A prominent example of
such a process is the M\"ossbauer effect \cite{migdal}.
It is an advantage that thermally isolated processes can be fast;
otherwise it is difficult to maintain thermal isolation, which is a
known problem of the original Carnot cycle \cite{callen}.

The work
\BEA
\label{dard}
W=W_1+W_2={\sum}_{k=2}^{n+1}(p^{[1]}_k-p^{[2]}_k)(\epsilon^{[1]}_k-\epsilon^{[2]}_k),
\EEA
is to be still maximized over $\{\epsilon^{[1]}_k\}_{k=2}^{n+1}$ and
$\{\epsilon^{[2]}_k\}_{k=2}^{n+1}$; see (\ref{buka}). $W$ is symmetric with respect to permutations
within $\{\epsilon^{[1]}_k\}_{k=2}^{n+1}$ and within $\{\epsilon^{[2]}_k\}_{k=2}^{n+1}$. We checked
numerically that this symmetry is not broken and hence the maximum of $W$ is reached for
\BEA
\label{kino}
\epsilon^{[\alpha]}\equiv\epsilon_2^{[\alpha]}=...=\epsilon_{n+1}^{[\alpha]}, ~~\alpha=1,2,
\EEA
i.e. all excited levels have the same energy. Thus the thermally
isolated pieces of the cycle consist, respectively, of sudden changes
$\epsilon^{[1]}\to\epsilon^{[2]}$ and
$\epsilon^{[2]}\to\epsilon^{[1]}$.

With new variables $e^{-\beta_\alpha \epsilon^{[\alpha]}}\equiv
u_\alpha$ we write the maximal work as
\BEA
\label{w}
W_{\rm max}[u_1,u_2]=\frac{(T_1\ln\frac{1}{u_1}-T_2\ln\frac{1}{u_2})(u_1-u_2)n}{[1+nu_1][1+nu_2]},
\EEA
where $u_1$ and $u_2$ are found from
\BEA
\label{det}
\partial_{u_1}W_{\rm max}[u_1,u_2]=\partial_{u_2}W_{\rm max}[u_1,u_2]=0.
\EEA
$u_1$ and $u_2$ depend on $T_2/T_1$ and on $n$. Noting (\ref{q1}) and
the result before (\ref{dard}) we obtain $Q_1={\rm
tr}(H_1(\rho_1-\rho_2))$ for the heat obtained from the high-temperature
bath. Using (\ref{kino}) we get from $\eta=W/Q_1$ and from (\ref{w}):
\BEA
\label{eta}
\eta=1-\left[T_2\ln{u_2}\right]\left/\left[T_1\ln{u_1}\right]\right..
\EEA
Note from (\ref{w}) that $W_{\rm max}[u_1,u_2]>0$ and $T_2<T_1$ imply
$1>\epsilon^{[2]}/\epsilon^{[1]}>T_2/T_1$. Hence (\ref{eta}) implies
$\eta\leq \eta_{\rm C}=1-T_2/T_1$, as expected.

Both $W_{\rm max}[u_1,u_2]$ and $\eta$ increase with $n$.
For $\ln [n]\gg 1$ we get asymptotically from (\ref{det}):
\BEA \label{u1u2}
u_1=\frac{(1-\theta)\ln[n]}{n}, ~~~~
u_2=\frac{\theta}{n\ln[n](1-\theta)},
\label{burda}
\EEA
where $\theta\equiv T_2/T_1$. This produces
\BEA
\label{kou}
W_{\rm max}[u_1,u_2]=(T_2-T_1)\ln[n]-{\cal O}({1}/{\ln[n]}), \\
\eta=\eta_{\rm C}-{\cal O}({1}/{\ln[n]}),~~ \eta_{\rm C}\equiv 1-T_2/T_1.
\label{go}
\EEA
The maximal work $W_{\rm max}[u_1,u_2]$ scales as $\ln[n]$,
since this is the ``effective number of particles" for the
engine. In the macroscopic limit $\ln[n]\gg 1$, the efficiency converges
to its maximal value $\eta_{\rm C}=1-T_2/T_1$; see (\ref{go}).

\textit{\textbf{The cycle time}} amounts to two times the relaxation
time $\tau_{\rm r}$ of the system with spectrum (\ref{kino}) and energy
gap $\epsilon\sim \ln[n]$; see (\ref{kino}, \ref{burda}). (Recall that
the thermally isolated stages of the cycle are very quick.) The
magnitude of $\tau_{\rm r}$ essentially depends on the scenario of
relaxation.

{\it First (specific) scenario.} We can assume that the Hamiltonian
(\ref{min}, \ref{kino}) of the heat engine is known. Then there exist
system-bath interaction scenarios that generally produce a non-Markovian
dynamics of the system and lead to $\tau_{\rm r}={\cal
O}(\sqrt{\ln[n]})$; see sections VI and VII of the Supplementary Material. Hence for
this type of relaxation the Carnot efficiency is achievable at a large
power ${\cal O}(\sqrt{\ln[n]})\gg 1$; see (\ref{kou}). However, in
these scenarios the system-bath interaction Hamiltonian (that governs
the relaxation) is special: it depends on the engine Hamiltonian
(\ref{min}, \ref{kino}).

{\it Second (realistic) scenario.} Assuming that the system-bath
interaction does not depend on the Hamiltonian (\ref{min},
\ref{kino}), we can estimate $\tau_{\rm r}$ within the weak-coupling,
Markov master equation approach that leads to $\tau_{\rm r}={\cal
O}(n)$; see section III of the Supplementary Material. For a qualitative
understanding of this situation, consider the relaxation as a random
walk in the energy space, e.g. in the second step of the cycle, where
the engine starts with almost unpopulated ground state, and it has to
achieve ground state probability $\approx 1$ after relaxation; see
(\ref{burda}). So, if every transition from one excited energy level
to another takes a finite time, one will need to perform in average
$n/2$ transitions before jumping to the ground state. Now note from
(\ref{go}) that the convergence of $\eta$ to $\eta_{\rm C}$ is
controlled by ${\cal O}(1/\ln[n])$: a small step towards $\eta_{\rm
C}$ will lead to a large increase in $\tau_{\rm r}$ nullifying the
power ${\cal O}(\ln [n]/n)$ for $n\gg 1$; see (\ref{kou}). Hence for
this type of relaxation the Carnot efficiency is not achievable at a
finite power.

The second relaxation scenario of the system with Hamiltonian
(\ref{min}, \ref{kino}) is similar to the known combinatorial
optimization problem: finding a marked item in an unstructured
database \cite{farhi} of $n+1$ items. This problem is mapped to
physics by associating each item to an eigenvector of a Hamiltonian
\cite{farhi}. The marked item relates to the lowest energy level $0$,
while all other (excited) eigenvalues of the Hamiltonian $\epsilon$
are equal. The resulting system has unknown eigenvectors of the
Hamiltonian, but known eigenenergies. Now the searching process can be
organized as a relaxation of the system from an excited state to a
low-temperature equilibrium state. This state is dominated by the
ground level due to a large $\epsilon$. Once the relaxation is over,
the unknown item (eigenvector) can be revealed by measuring the energy
\cite{foxnote}.

For classical algorithms the search time of this problem scales as
${\cal O}(n)$ for $n\gg 1$ \cite{farhi}. It is thus not much better
than going over all possible candidates for the solution, a typical
situation of a computationally complex problem. For quantum algorithms
(Grover's search) the search time scales as ${\cal O}(\sqrt{n})$
\cite{farhi}. This is still not suitable for our purposes, since it
nullifies the power for $\ln[n]\gg 1$.

\textit{\textbf{ Sub-optimal engine.}} Within the second (realistic)
relaxation scenario, we shall modify the optimal engine so that the
power is finite, but both the work and efficiency are still large. We
are guided by the analogy between the relaxation of the Hamiltonian
(\ref{min}, \ref{kino}) under the second scenario and the Levinthal's
paradox from protein physics \cite{zwan}. In fact, (\ref{min},
\ref{kino}) is the simplest model employed for illustrating the
paradox; see \cite{zwan,book} and section V of the Supplementary Material. Here the
ground state refers to the unique folded (native) state. To ensure its
stability, it is separated by a large gap from excited (free) energy
levels. The essence of the paradox is that assuming
many {\it equivalent} unfolded (excited) states, the relaxation time
to the native state is unrealistically long. Recall that the
states $\rho_1$ and $\rho_2$ of the optimal engine refer respectively
to unfolded and folded states of the protein model. Indeed
$nu_\alpha/(1+nu_\alpha)$ ($\alpha=1,2$) is the overall probability of
the excited levels; see (\ref{kino}). Hence for $\ln[n]\gg 1$ the
ground state (excited levels) dominates in $\rho_2$ ($\rho_1$); see
(\ref{burda}).

The resolution of the paradox is to be sought via resolving the
degeneracy of excited levels: if there are energy differences, some
(unfavorable) transitions will not be made shortening the relaxation
time \cite{zwan,book}. In resolving the energy degeneracy we follow the
simplest model proposed in \cite{zwan}.

The model has $N\gg 1$ degrees of freedom $\{\sigma_i\}_{i=1}^n$; each
one can be in $\zeta+1$ states: $\sigma_i=0,...,\zeta$. Whenever
$\sigma_i=0$ for all $i$'s, the model protein is in the folded (ground)
state with energy zero \cite{zwan,book}. The ground state has zero
energy. Excited states with $s\geq 1$ have energy $\epsilon+\delta
s$, where $\epsilon>0$ and $s$ is the number of (misfolded) degrees of
freedom with $\sigma_i\not=0$. $\delta>0$ is the parameter that
(partially) resolves the degeneracy of excited states; we revert to
the previous, work-optimal, model for $\delta\to 0$.
For {\it different} eigenvalues of the Hamiltonian $H_\alpha$ we have
\BEA
\left\{\, (1-\delta_{\rm Kr}[s,0]\,)\,(\epsilon^{[\alpha]}+s
\delta^{[\alpha]})\right\}_{s=0}^N, ~~ \alpha=1,2,
\label{nova}
\EEA
where $\delta_{\rm Kr}[s,0]$ is the Kronecker delta, and
where each energy $\epsilon^{[\alpha]}+s \delta^{[\alpha]}$ is
degenerate $\frac{\zeta^s N!}{s!(N-s)!}$ times; thus the total number
of energy levels is $(1+\zeta)^N$.

Given (\ref{nova}), the cycle consists of two isothermal and two
thermally isolated pieces with sudden changes $(\delta^{[1]},
\epsilon^{[1]})\to (\delta^{[2]}, \epsilon^{[2]})\to (\delta^{[1]},
\epsilon^{[1]})$; see (\ref{0}--\ref{dard}). Below we shall also
assume
\BEA
\label{bovi}
\beta_1\delta^{[1]}=\beta_2\delta^{[2]},
\EEA
because this makes the sub-optimal engine structurally very similar to
the optimal one. Now the work $W=W_1+W_2$ is
calculated from (\ref{w1}, \ref{w2}, \ref{q1}, \ref{nova}, \ref{bovi}):
\BEA
\label{wi}
&& W[v_1,v_2;K]=\frac{m(\Delta\epsilon + \frac{KN\Delta\delta}{1+K})(v_1-v_2)}{(1+mv_1)(1+mv_2)}, \\
&& \Delta\epsilon=\epsilon^{[1]}-\epsilon^{[2]} = T_1\ln [{1}/{v_1}] - T_2\ln[{1}/{v_2}],\\
&& \Delta\delta=\delta^{[1]}-\delta^{[2]}=(T_1-T_2)\ln [\zeta/K].
\label{wiwi}
\EEA
where $K=\zeta e^{-\beta_1\delta^{[1]}}$, $m=(1+K)^N$, and where
$v_\alpha\equiv e^{-\beta_\alpha \epsilon^{[\alpha]}}$ ($\alpha=1,2$) are determined
from maximizing (\ref{wi}). Note the analogy between (\ref{w}) and
(\ref{wi}), with $m$ being an analogue of $n$; they get equal for
$\delta\to 0$. Note that in (\ref{wi}) we neglected factor ${\cal O}(\frac{1}{m})$
assuming that $m\gg 1$.

\begin{table}
\begin{center}
\tabcolsep0.03in \arrayrulewidth0.5pt
\renewcommand{\arraystretch}{1.6}
\caption{ Parameters of the sub-optimal engine: work $W$, efficiency
$\eta$ and the cycle time $2\tau_{\rm r}$; see
(\ref{wi}--\ref{rela}). $W_{\rm max}$ is the maximal work extracted
  for the optimal engine at a vanishing power; see (\ref{w},
  \ref{det}). For the sub-optimal engine: $K=\zeta
  e^{-\beta_1\delta^{[1]}}$, $N=140$, $\zeta=4$, $T_1=1$,
  $T_2=1/2$. Carnot and Curzon-Ahlborn efficiencies are, respectively,
  $\eta_{\rm C}=1/2$ and $0.5858\, \eta_{\rm C}$. Also,
  $p^{[\alpha]}_1=[1+(1+K)^N e^{-\beta_\alpha
    \epsilon^{[\alpha]}}]^{-1}$ ($\alpha=1,2$) are the
  ground-state probabilities of $\rho_\alpha\propto e^{-\beta_\alpha
    H_\alpha}$; see (\ref{nova}).  }
\begin{tabular}{|c||c|c|c|c|c|c|}
\hline
  $K$     & $\tau_{\rm r}$  & $W/W_{\rm max}$ & $W$ & $\eta/\eta_{\rm C}$ & $p^{[1]}_1$ & $p^{[2]}_1$ \\
\hline\hline
 $0.1$    & $4.45\times 10^{-5}$ {\rm s} & $0.2267$  &  $23.52$ & $0.8751$ & $0.0392$ & $0.9808$ \\
\hline
 $0.2$    & $4.35$ {\rm s}  & $0.3884$  & $40.3$  & $0.9110$ & $0.0237$ & $0.9883$ \\
\hline
 $0.24$    & $357$ {\rm s}  & $0.4393$  & $45.58$  & $0.9181$ & $0.0210$ & $0.9896$ \\
\hline
\end{tabular}
\end{center}
\end{table}

Likewise, we get for the efficiency [cf. (\ref{eta})]:
\BEA
\label{etaa}
\eta=1-\frac{T_2}{T_1}\times\frac{\ln\frac{1}{v_2}
+\frac{NK\ln(\zeta/K)}{1+K} }{\ln\frac{1}{v_1} +\frac{NK\ln(\zeta/K)}{1+K} }.
\EEA
For this model \cite{zwan} assumes a
local Markov relaxation dynamics, where each degree of freedom makes a
transition $\sigma_i\to\sigma_i\pm 1$ in $10^{-9}$ seconds; this value
is chosen conventionally to fit experimental magnitudes for the
elementary dynamic step \cite{zwan}. The model has a single relaxation
time \cite{zwan} that is easily reproduced in the general
master-equation framework (see section IV of the Supplementary Material):
\BEA
\tau_{\rm r}=10^{-9}(1+K)^N/(NK)~ {\rm seconds},
\label{rela}
\EEA
where the factor $N$ is due to the $N$-times degenerate first excited level.

For $\delta^{[\alpha]}\to 0$ ($\alpha=1,2$), where the excited energy
levels become degenerate, $\tau_{\rm r}\propto (1+\zeta)^N$ scales
linearly over the number of energy levels, as expected. When
$\delta^{[\alpha]}$ are not zero, $\tau_{\rm r}$ can be of order of $1$
second for $N\sim 100$, because $1+K$ is close to $1$. However, for the
macroscopic situation ($N\sim 10^{23}$) $\tau_{\rm r}$ is still
huge. In this sense, the model is incomplete, but still
useful for analyzing the mesoscopic situation $N\sim 100$ that is
relevant for the protein folding problem \cite{book}.

Table~I illustrates the characteristics of the sub-optimal engine and
compares them with those of the optimal one. Reasonable cycle times
can coexist with a finite fraction ($\sim 40\%$) of the maximal work
and with sizable efficiencies ($\sim 90\%$ of the maximal value) that
are larger than the Curzon-Ahlborn value. Hence, albeit within
the second (realistic) scenario it is impossible to approach the
maximal efficiency as close as desired, reasonably large efficiencies
at a finite (or even large) power are possible. These results resemble
the power-efficiency trade-off (see \cite{tarm5}), but they are more
complicated, since they involve work, efficiency and power.

K.V.H. is supported by the Spanish project FIS2010-14830. S.G.G. is
supported by Grant NSC 101-2811-M-001-156.


\appendix

\section*{Supplementary Material}

This material consists of seven sections. Almost all of them can be
read independently.

Section \ref{geno} clarifies the finite power condition for the
generalized Carnot cycle and it compares it with the usual Carnot
cycle.

Section \ref{works} relates
together two definitions of (thermodynamical) work.

Section \ref{rela} estimates the relaxation time of the optimal engine
within the master-equation framework. Since this is one of the main
points of the present work, we dwell on it in detail and spell out all
(hidden) assumptions necessary for its derivation. Section \ref{subo}
presents a similar estimation for the sub-optimal engine.

In section \ref{levin} we discuss the current status of the
Levinthal's paradox within the protein folding theory.

Section \ref{gegemon} discusses an example of the quantum relaxation
scenario, where the system-bath interaction is engineered, i.e., it
correlates with the system Hamiltonian. We show that the relaxation
time of the optimal engine within this scenario is short. Due to this
fact, the Carnot efficiency can be reached at a large power. Finally,
in section \ref{enta} we show that this shortening of the relaxation
time is not related to the extensive usage of resources such as
quantum entanglement.

\section{Power for the (generalized) Carnot cycle}
\label{geno}

Here we discuss in some details the power of the generalized Carnot
cycle and compare this situation with the usual Carnot cycle.

Recall from the main text that the generalized Carnot cycle consists
of four pieces: two of them are thermally isolated that can proceed
very fast. The rate-limiting steps are the two pieces with free
relaxation, since their duration is bound by the relaxation time.

To achieve a cyclic process within the exponential relaxation with the
relaxation time $\tau_{\rm r}$, the cycle time $\tau$ should be larger
than $\tau_{\rm r}$. For each cycle the deviation of the
post-relaxation state from the exact equilibrium (Gibbsian) state
will be of order $e^{-\tau/\tau_{\rm r}}$.  Thus if
the ratio $\tau/\tau_{\rm r}$ is simply large, but finite, one can
perform roughly $\sim e^{\tau/\tau_{\rm r}}\gg 1$ number of cycles at
a finite power, before deviations from cyclicity would accumulate and
the machine will need resetting.

The above situation does differ from the power consideration of usual
(reversible) thermodynamic cycles, e.g., the Carnot cycle
\cite{landau,sekimoto,sekimoto_sasa}. There the external fields
driving the machine through various stages have to be much slower than
the relaxation to the momentary equilibrium. The latter means that the
machine is described by its equilibrium Gibbs distribution with
time-dependent parameters. In particular, the condition of momentary
equilibrium for the working medium is necessary for the Carnot cycle
to reach the Carnot efficiency \cite{landau}.

The precise meaning of the external fields being slow is important
here. If $\tau_{\rm F}$ is the characteristic time of the fields, then
the deviations from the momentary equilibrium are of order ${\cal
O}[\frac{\tau_{\rm r}}{\tau_{\rm F}}]$
\cite{landau,sekimoto,sekimoto_sasa}. This fact is rather general and
does not depend on details of the system and of the studied process,
e.g., it does not depend whether the process is thermally isolated or
adiabatic.
In particular, it is this deviation of the state from the momentary equilibrium that
brings in the entropy production (or work dissipation) of order of
${\cal O}[\left(\frac{\tau_{\rm r}}{\tau_{\rm F}}\right)^2]$
\cite{landau,sekimoto,sekimoto_sasa}.

Thus performing the reversible Carnot cycle with (approximately) the
Carnot efficiency means keeping the ratio $\frac{\tau_{\rm
    r}}{\tau_{\rm F}}$ very small.

Here are the differences between the Carnot cycle and our situation:

\begin{itemize}

\item In our case we do not require the machine to be close to
  its momentary equilibrium state during the whole process. It
  suffices that the machine gets enough time to relax to its final
  equilibrium.

\item A small, but finite $\frac{\tau_{\rm r}}{\tau_{\rm F}}$ for
  the Carnot cycle situation means that deviations from the momentary
  equilibrium are visible already within one cycle. In contrast, a
  small, but finite $\frac{\tau_{\rm r}}{\tau}$ for our situation
  means that we can perform an exponentially large number of cycles
  before deviations from the cyclicity will be sizable. Here is a
  numerical example. Assume that $\frac{\tau_{\rm
      r}}{\tau}=\frac{\tau_{\rm r}}{\tau_{\rm F}}={1}/{20}$. For
  the standard Carnot cycle already within one cycle the deviation
  from the momentary equilibrium will amount to $0.05$. In our
  situation the same amount $e^{-3}=0.0498$ of deviation from the
  cyclicity will come after $e^{17}=2.4\times 10^7$ cycles. This is a
  large number, especially taking into account that no realistic
  machine is supposed to work indefinitely long. Such machines do need
  resetting or repairing. The point is that our machine can perform
  {\it many} cycles at a finite power before any resetting is
  necessary.

\end{itemize}

\section{Clarification of the concept of work as used in our situation.}
\label{works}

Let a system interact with a source of work only (thermally isolated
process). This means that the system Hamiltonian $H[\alpha(t)]$ is a
function of a (classical) parameter $\alpha(t)$. The work done on the
system per unit of time equals to "force" times "displacement" and
averaged over state of the system, as represented by a time-dependent
density matrix $\rho(t)$:
\BEA
\label{ekler}
\frac{\d W}{\d t}= \frac{\d \alpha(t)}{\d t} {\rm tr}\left(\frac{\partial H}{\partial \alpha}
\rho(t)  \right).
\EEA
The same formula applies in the classical situation, where ${\rm tr}$
means integration over the phase-space (the space of coordinates and
momenta), while $\rho$ becomes the phase-space probability density. Now
one can use the equations of motion for the density matrix,
$i\hbar\frac{\d \rho}{\d t}=H[\alpha(t)]\rho(t)-\rho(t)H[\alpha(t)]$ (in
the classical situation this becomes the Liouville equation), to show
from (\ref{ekler}) that the total work equals to the change of average
energy
\BEA
W(\tau)\hspace{-0.4mm}=\hspace{-1.3mm}\int_0^\tau \hspace{-1.5mm}\d t\, \frac{\d W}{\d t} =
{\rm tr}\left( H[\alpha(\tau)] \rho(\tau) -H[\alpha(0)] \rho(0) \right)\hspace{-0.5mm}.~~~
\EEA

\section{Relaxation time of the optimal engine via master equation}
\label{rela}

{\bf 1.} Consider a system with $n\gg 1$ degenerate (excited) levels
with energy $\varepsilon>0$, and a single ground state with energy
$0$. In this section we shall estimate the relaxation time of this
system within the Markov master-equation framework. We start with
simplifying assumptions for showing the origin of a long relaxation
time in this system. We then demonstrate the result at more general
level.

Let $\{p_i\}_{i=0}^{n}$ be the probability of energy
levels. The master equation reads \BEA
\label{eq}
\dot{p}_0={\sum}_{i=1}^{n} w_{0i}p_i-p_0{\sum}_{i=1}^{n} w_{i0},
\EEA
where $w_{0i}$ is rate of the transition $i\to 0$. Since all energy
levels besides the lowest one have the same energy $\varepsilon$,
the detailed balance condition reads (which reflects
the fact that the bath is in equilibrium at temperature $1/\beta$):
\BEA \label{db}
w_{0i}e^{-\beta\varepsilon}=w_{i0}.
\EEA

{\bf 2.}
Let us now assume that in (\ref{eq}) all the excited energy levels are
equivalent, and hence
\begin{eqnarray}
  \label{orinoco}
  w_{0i}=w_{01}~~~ {\rm and}~~~ w_{i0}=w_{10},
\end{eqnarray}
do not depend on $i$' (this assumption is relaxed at the end of the section):
\BEA
\label{be}
\dot{p}_0=-  w_{01}p_0[n e^{-\beta \varepsilon}+1]+w_{01},
\EEA
meaning that the relaxation time $\tau_{\rm r}$ for $p_0(t)$
to converge exponentially to its equilibrium value
\BEA
\label{equ}
p^{\rm [eq]}_0 = 1/(1+n e^{-\beta\varepsilon})
\EEA
reads
\BEA
\label{res}
{1}/{\tau_{\rm r} }=w_{01}[n e^{-\beta \varepsilon}+1].
\EEA

A popular choice for the rate is given by the transition state theory
\cite{review}:
\begin{eqnarray}
  \label{eq1}
w_{01}=\frac{\kappa}{n+1}\,e^{-\beta (\varepsilon^*-\varepsilon) },
\end{eqnarray}
where $\varepsilon^*>\varepsilon$ is the transition state energy: once
the system gets at that state, it has equal probability to move to any
state; hence the factor $\frac{1}{n+1}$ in (\ref{eq1}).  Here $\kappa$
does not depend on $n$; it is determined by the the energy landscape
in the vicinity of the transition state and the excited state.

In our situation
\begin{eqnarray}
  \label{e}
e^{-\beta\varepsilon}={\cal O}(\ln[n]/n)~~~{\rm or}~~~
e^{-\beta\varepsilon}={\cal O}(\frac{1}{n\ln[n]}).
\end{eqnarray}
Eqs.~(\ref{eq1}, \ref{res}) then imply that the
relaxation time $\tau_{\rm r}$ is roughly (neglecting logarithmic
factors)
\begin{eqnarray}
  \label{eq2}
  {\tau_{\rm r} }={\cal O}(n)\gg 1.
\end{eqnarray}
This conclusion is then based on two physical aspects: first that
there are many states with the same energy [the factor
$\frac{\kappa}{n+1}$ in (\ref{eq1})]. Second is that the energy gap is
large; see (\ref{e}).

{\bf 3.}
For more general (than the transition state theory) choices of
$w_{01}$ we still shall obtain the same result if we require that for any
excited state energy $\varepsilon>0$, (\ref{be}) produces a
well-defined and finite limit for $n\to\infty$. This request is based
on two hidden assumptions: {\it i)} the master equation is derived
within the weak-coupling assumption, hence it cannot contain very fast
(in the limit $n\to\infty$) characteristic times; {\it ii)} the
limit $n\to \infty$ can be taken independently from $\varepsilon$.
Now we naturally get that $w_{01}$ and $w_{10}$
have an overall dependence ${\cal O}(\frac{1}{n})$:
\BEA
\label{kran}
w_{01}= \hat{w}_{01}(\varepsilon)/n, ~~
w_{10}= \hat{w}_{10}(\varepsilon)/n,
\EEA
where
$\hat{w}_{10}(\varepsilon)$ and $\hat{w}_{01}(\varepsilon)$ can depend
on energy $\varepsilon$, but they do not depend on $n$ directly
[cf. (\ref{eq1})].  Then (\ref{res}) assumes a finite limit for $n\gg
1$.  We now get from (\ref{kran}, \ref{e}):
\begin{eqnarray}
  {1}/{\tau_{\rm r} }=\hat{w}_{01}(\varepsilon)
\times {\cal O}(\frac{1}{n}).
\end{eqnarray}
It is natural to assume that for $\varepsilon\to\infty$,
$\hat{w}_{01}(\varepsilon)$ stays at least bounded [cf. (\ref{eq1},
\ref{e})], and then we are back to (\ref{eq2}).  There is an important
relaxation scenario (going back to Arrhenius and improved by Kramers)
where $\hat{w}_{01}$ does not depend on $\varepsilon$ (provided that
$\varepsilon>0$ is sufficiently large); it is given by
$\hat{w}_{01}\propto e^{-\beta V}$, where $V>0$ is the barrier height
\cite{review}.

Thus we note that the conclusion (\ref{eq2}) on long characteristic
times is not completely straightforward and|if taken out of the usual
relaxation theories, e.g. the transition state theory or the Arrhenius
theory|it requires several hidden assumptions. In section V of this
supplementary material we show that (\ref{eq2}) is violated, and the
relaxation time can be much shorter, if allow the system-bath
interaction to depend on the system features.

{\bf 4.}
Finally, let us return to (\ref{orinoco}) and show that this assumption
can be relaxed without changing our main conclusions. For a
sufficiently large energy gap $\varepsilon>0$ between the ground state
and the excited state, we can apply the adiabatic approximation
meaning that the excited levels probability equilibrate between
themselves much quicker than the ground state level probability. Hence
they all get into the same value:
\begin{eqnarray}
  \label{eq:5}
  p_i(t)=(1-p_0(t))/n,
\end{eqnarray}
before $p_0(t)$ start to change appreciably.  Employing (\ref{eq:5})
in (\ref{eq}) we revert to (\ref{eq}), where now instead of $w_{01}$
we should employ $\frac{1}{n}\sum_{i=1}^n w_{0i}$. Note that
(\ref{eq:5}) is especially plausible in our situation, since we also
start the relaxation process from equilibrium states at a temperature
different from the bath temperature. At such an initial state the
probabilities of the excited levels are equal.

{\bf 5.}
To go beyond the adiabatic approximation we will now consider the whole
master equation. Let us first note that the detailed balance conditions
(\ref{db}) for the transition rates $w_{ij}$ between excited levels lead
to $w_{i\neq j}=w_{ji}$; and write down the rest of the master equation (\ref{eq})
\BEA
\dot{p}_i=\sum_{j\neq i}w_{ij}p_j-\sum_{j\neq i}w_{ji}p_i-w_{0i}p_i+w_{i0}p_0
\EEA
in the following form:
\BEA \label{me1}
\dot{p}_i=\sum_{j}\bar{w}_{ij}p_j-w_{0i}p_i+e^{-\beta\varepsilon}w_{0i}p_0.
\EEA
Where the matrix $\bar{w}$ is defined as: $\bar{w}_{ii}=-\sum_{j\neq i}w_{ij}$,
$\bar{w}_{i\neq j}=w_{ij}$; and is a symmetric matrix satisfying $\sum_iw_{ij}=0$
for $\forall j$ and can be shown to be non-positive.

We now form the quantities $Q_i=p_i-e^{-\beta\varepsilon}p_0$, for
which the master equation (\ref{eq}, \ref{me1}) produces
\BEA \label{me2} \dot{Q}_i=\sum_j
\bar{w}_{ij}Q_j-w_{0i}Q_i-e^{-\beta\varepsilon}\sum_j w_{0j}Q_j.  \EEA
Introducing the positive and symmetric matrix
$W_{ij}=-\bar{w}_{ij}+\delta_{ij}w_{0i}$ and bra-ket notation for
vectors, we rewrite (\ref{me2}) as
\BEA
\frac{d}{dt}|Q\rangle=-\left(W+e^{-\beta\varepsilon}|1\rangle\langle
  w|\right)|Q\rangle,
\EEA
where $\langle1|=\left(1,...,1\right)$ and
$\langle w|=\left(w_{01},...,w_{0n}\right)$.
In the equilibrium state $Q_i= 0$.

Now the relaxation time of the system can be estimated via the minimal
eigenvalue $\omega_{\rm min}$ of the matrix
$W+e^{-\beta\varepsilon}|1\rangle\langle w|$.
Below, we show numerically that in the asymptotic limit of $n\to\infty$,
\BEA
\label{eigen}
&& \omega_{\rm min}=\left(1+n e^{-\beta\varepsilon}\right)\frac{\sum_i
  w_{0i}}{n} +\frac{f_n}{n^2}, \\
&& f_n={\cal O}\left(1\right),
\label{eigeno}
\EEA
where $f_n$ does not depend on $n$ for a sufficiently large $n$.
Hence for $n\gg 1$ we neglect the last term in
(\ref{eigen}) and obtain for the relaxation time:
\BEA \frac{1}{\tau_{\rm
    r}}=\left(1+n e^{-\beta\varepsilon}\right)\frac{\sum_i w_{0i}}{n},
\label{boru}
\EEA
confirming that conditions (\ref{kran}, \ref{e}) imply the relaxation
time to grow proportionally with $n$.  Eq.~(\ref{boru}) is the same
result as was obtained above via the adiabatic approximation. Hence
this approximation holds up to the second-order term of the asymptotic
expansion of $\omega_{\rm min}$ in terms of $n$.

We now demonstrate (\ref{eigen}) numerically. As it
suggests, it holds also when $e^{-\beta\varepsilon}$ depends on $n$ as
in formulas (\ref{e}). We present our numerical evidence for an
illustrative case of $e^{-\beta\varepsilon}={\rm const}/n$.  Note,
that all $w_{0i}$s and $w_{ij}$s scale as $1/n$ as in (\ref{kran}). To
prove (\ref{eigen}) we calculated $f_n$ for random collections of $n
w_{0i}$s and $n w_{ij}$s with various probability distributions; see
Table~\ref{tab} for illustration.
\begin{table}
\begin{center}
\tabcolsep0.03in \arrayrulewidth0.5pt
\renewcommand{\arraystretch}{1.6}
\caption{ Statistics for $f_n$ -- the mean $\langle f_n\rangle$, and
  standard deviation $\left\langle (f_n-\langle
    f_n\rangle)^2\right\rangle^{1/2}$ -- are shown for different
  values of $n$; see (\ref{eigen}, \ref{eigeno}).
  It is seen that both depend on $n$ very weakly. \\
  The Boltzmann weight $e^{-\beta\varepsilon}=0.2/n$. The positive
  transition probabilities $\hat{w}_{ij}=nw_{ij}=nw_{ji}$ and
  $\hat{w}_{0k}=nw_{0k}$ are random variables.  They are all
  independent from each other. $\hat{w}_{ij}$ and
  $\arcsin[\hat{w}_{0k}]$ are uniformly distributed in the interval
  $(0,1)$. The $\arcsin$ function for generating $\hat{w}_{0i}$ is
  chosen with no special reason (other functions were tried as well
  with similar results), it is just taken to make distributions
  different.  }
\begin{tabular}{|c|c|c|}
  \hline
  $n$ & $\langle f_n\rangle$ & $\left\langle (f_n-\langle f_n\rangle)^2\right\rangle^{1/2}$ \\
  \hline\hline
  \quad 130 \;\;\quad & \quad\quad\ 0.147179 \quad\quad\quad & \quad\quad\ 0.009826 \quad\quad\quad \\
  \hline
  150 & 0.149095 & 0.010866 \\
  \hline
  175 & 0.146339 & 0.009440 \\
  \hline
  200 & 0.147149 & 0.009600 \\
  \hline
  230 & 0.147958 & 0.008707 \\
  \hline
\end{tabular}
\end{center}
\label{tab}
\end{table}

\section{Relaxation time of sub-optimal engines via adiabatic
  approximation of the master equation}
\label{subo}

We return to (\ref{eq}), but we do not assume anymore that the excited
levels $\varepsilon_1,..., \varepsilon_n$ have the same energy. But we still
assume that the gap $\varepsilon_1>0$ between the zero-energy ground state
and the first excited state is the largest energy parameter in the
system, because we want to ensure that the equilibrium ground-state
probability is close to $1$. Thus we may apply to (\ref{eq}) the
adiabatic approximation assuming that on those times where $p_0$
changes, the excited-state probabilities $p_i$ already equilibrated:
\BEA
\label{krot}
p_i(t)=(1-p_0(t)) e^{-\beta\varepsilon_i}\left/ {\sum}_{i=1}^n e^{-\beta\varepsilon_i}\right. .
\EEA
Hence the relaxation time $\tau_{\rm r}$ of $p_0$
deduced from (\ref{eq}, \ref{krot}) reads
\BEA
\label{je}
\frac{1}{\tau_{\rm r} }= \frac{{\sum}_{i=1}^n w_{0i} e^{-\beta\varepsilon_i}}{ {\sum}_{i=1}^n e^{-\beta\varepsilon_i}}
+ {\sum}_{i=1}^n e^{-\beta\varepsilon_i} w_{0i},
\EEA
where the employed the detailed balance condition.

Eq.~(\ref{je}) reproduces the relaxation time of the Zwanzig model
\cite{zwan} [given by (27) of the main text], if we employ the energy
spectrum (21) [of the main text] and note that the transition
probabilities from the first excited ($\zeta N$-degenerate) energy
level are constants, $w_{0k}= 10^{9}$ for $k=1,...,\zeta N$, while no
transitions (to the ground state) is possible from other excited
states: $w_{0l}=0$ for $l=\zeta N+1,...,n$.  Here $10^9$ is the
characteristic microscopic scale \cite{zwan}.  Putting these into
(\ref{je}) we get that the first term in the right-hand-side of
(\ref{je}) reproduces (27) of the main text. The second term is
negligible, if $e^{-\beta\varepsilon_1}$ is sufficiently small.

Note that according to (27) of the main text, the relaxation time of
the Zwanzig model is still unacceptably large, if $N\gg 1$. It is
possible to get rid of this restriction, but doing so is not useful for
the engine functioning.

\section{A short reminder on the protein folding theory in the
  context of the Levinthal's paradox}
\label{levin}

The purpose of this section is to introduce the reader into some of
the current ideas in protein folding theory. In particular, this
should prevent confusions on how specifically we employ the
Levinthal's paradox in our study.

{\bf 1.} Early experiments have shown that proteins can fold
(i.e. reach the native, functional state) in a reasonably short time;
see \cite{rev_prot}. Moreover, they do so spontaneously (without
external guidance) and starting from different initial
conditions. These experiments created what is known to be the modern
thermodynamic paradigm on the protein folding: the native state
corresponds to the (relatively) unique global minimum of free energy
\cite{rev_prot}.

{\bf 2.} It however still remained unclear how specifically proteins
fold, i.e. what is their kinetics. Levinthal assumed that all
unfolded states (conformations) are more or less equivalent
\cite{levin}. Hence during the relaxation to the folded state all
possible conformations are tried out to find the energetically most
favorable one. This will take an enormous amount of time, because for
a (hypothetical) small protein with $100$ residues, the number of
possible unfolded states would be about $3^{100}$ \cite{levin}. Since
this conclusion is clearly unsupportable (hence the Levinthal's
paradox), there should be some structure in the set of unfolded states
that makes them non-equivalent.

{\bf 3.} One (by now classical) view suggests that in the course of
its relaxation the protein passes through a unique path of partially
unfolded intermediate states \cite{baldwin}. They allow stepwise
folding, drastically reducing the scale of conformational search. In
that view|which was supported by experiments on sufficiently long
proteins \cite{baldwin}|the protein folding problem is reduced to
relaxation in a finite number of states \cite{rev_prot}. Still it was
unclear how the protein reaches one of those intermediates, since now
the Levinthal's paradox can be reformulated with respect to partial
relaxation.

{\bf 4.} In addition to the latter objection, experiments on short
proteins have shown that folding intermediates are absent. Hence a new
view emerged that explains the protein folding kinetics as taking
place on a funnel (free) energy landscape, where different unfolded
states have different (free) energies and it is this difference that
drives the protein towards the minimum (free) energy state
\cite{zwan,wol}. An example of the above scenario is provided by the
Zwanzig's model \cite{zwan}, as reviewed above.

Still for realistic proteins the Zwanzig's model is clearly
oversimplified, e.g., it does not include conformational entropy,
disorder, residue sequence, {\it etc}. Thus for a deeper understanding of
the protein folding one should go to more realistic models \cite{wol},
which however share the two main points of the the Zwanzig's model:
{\it i)} unfolded states have different (free) energies;
{\it ii)} the folding time can be made finite via fine-tuning
only for sufficiently short proteins having $100-200$ coarse-grained
degrees of freedom. It is argued that longer proteins will fold
hierarchically, i.e. first certain domains will fold independently
from each other (these domains thus play the role of folding
intermediates), and only after that the protein will relax
globally. The existence of fine-tuning is explained via evolution, a
notorious solver of difficult problems in biology \cite{rev_prot}.

\section{Fast relaxation scenario}
\label{gegemon}

We shall now study a quantum model of relaxation that achieves a fast
relaxation of the work-optimal engine at the cost of introducing
specific system-bath interaction Hamiltonian.

\subsection{System-bath interaction}

Consider a system E with $n$ excited energy levels and one lowest energy
(ground state) whose energy we set to zero. All $n$ excited levels have
the same energy $\epsilon$.

The initial density matrix of E is Gibbsian at temperature $T_0=1/\beta_0$:
\BEA
\rho\propto e^{-\beta_0 H}
=r \P_0 +\frac{1-r}{n}\P_\epsilon, \\
r =\frac{1}{1+n e^{-\beta_0\epsilon}},
\EEA
where $\P_0=|0\rangle\langle 0|$ is the projector on the ground state,
and $\P_\epsilon$ is the projectors on the $n$-dimensional eigen-space
of $\rho$ with eigenvalue $\epsilon$.

Now $E$ interacts with an external thermal bath at temperature
$T=1/\beta$, so that the density matrix of E converges in time to
$\rho_{eq}\propto e^{-\beta H}$. We shall design a concrete model
for this interaction and estimate the relaxation time.

We assume that the bath consists of a large number of independent
particles prepared in identical (thermal states). E interacts with one
particle, then with the second one {\it etc}. Since the particles are
independent, it will suffice to consider the interaction of E with the
first particle B only.

We assume that the bath particle B has (among other energies) energy
levels $E$ and $E+\eps$. The degeneracies of these levels are $n_{E}$
and $n_{E+\eps}$, respectively.  The initial (before interacting with E)
equilibrium density matrix of B reads
\BEA
\label{fru}
\sigma =e^{-\beta H_{\rm B}}/Z=
\widetilde{\sigma} +r_E \, \Pi_E +r_{\epsilon+E}\, \Pi_{\epsilon+E}, \\
r_{E}={e^{-\beta E}}/{Z}, \quad Z={\sum}_E n_{E} e^{-\beta E},
\label{blob}
\EEA
where $r_E$ and $r_{E+\eps}$ are the Boltzmann weights for
the energy levels $E$ and $E+\eps$, respectively, the summation in
(\ref{blob}) is taken over all energy levels of B.
$\Pi_E$ and $\Pi_{\eps+E} $ are the projectors on the
corresponding sub-spaces,
\BEA
{\rm tr}\,\Pi_E=n_{E},    \qquad
{\rm tr}\,\Pi_{\eps+E}=n_{E+\eps},
\EEA
and where $\widetilde{\sigma}$ in (\ref{fru}) is the remainder of $\sigma$.

It is assumed that the unitary operator ${\cal V}$ responsible for the
interaction operates within the sub-space with the projector
$\P_\eps\otimes \Pi_E + \P_0\otimes \Pi_{E+\eps}$ (this
sub-space has energy $E+\eps$), i.e.,
\BEA
[\, {\cal V},\P_\eps\otimes \Pi_E + \P_0\otimes \Pi_{E+\eps}\, ]=0.
\label{brams}
\EEA
On the remainder of the overall Hilbert space (of E+B) ${\cal V}$ acts
as unit operator. Thus, ${\cal V}$ commutes with the Hamiltonian of E+B.
Hence no additional energy (work) is needed for switching the E-B
interaction on and off. In that respect ${\cal V}$ resembles the
weak-coupling, though by itself it need not be weak, i.e. it need not be
smaller than the Hamiltonian of E+B.

Then the post-interaction density matrix $\rho'$ of E reads
\BEA
\rho'&=& {\rm tr}_{\rm B}{\cal V} \rho\otimes\sigma{\cal V}^\dagger \nonumber\\
     &=& \rho
-\left(r\,r_{E+\eps}-r_E\,\frac{1-r}{n}\right)\times \nonumber\\
&& [n_{E+\eps}\P_0
-{\rm tr}_{\rm B}{\cal V}\P_0\otimes \Pi_{E+\eps} {\cal V}^\dagger].
\label{dali}
\EEA
It will suffice to keep track of the lowest energy-level
occupation $\langle 0|\rho'|0\rangle\equiv r'$ only:
\begin{gather}
\label{cuba}
r'-r=-A\left [ r- r_{\rm eq} \right], \quad r_{\rm eq}\equiv \frac{1}{1+n e^{-\beta\eps}},\\
\label{gvadalajara}
A\equiv \frac{r_E}{r_{\rm eq}n}
\left [n_{E+\eps}
-\langle 0|\,(\, {\rm tr}_{\rm B}{\cal V}\P_0\otimes \Pi_{E+\eps} {\cal V}^\dagger\,)|0\rangle \right],
\end{gather}
where $r_{\rm eq}$ is the equilibrium value of $r$.
Using (\ref{blob}) one can show that $A\leq A_{\rm max}\leq 1$:
after the interaction E gets closer to its equilibrium state;
see (\ref{cuba}).

Now (\ref{dali}) serves as the initial state of E for a similar
interaction with the second bath particle that initially has the same
state $\sigma$ as in (\ref{fru}). We get for all subsequent interactions
[we revert from (\ref{cuba1}) to (\ref{cuba}) for $m=1$]:
\BEA
\label{cuba1}
r^{[m]}-r_{\rm eq}=(1-A)^m\left [ r- r_{\rm eq} \right],
\EEA
It is seen that (\ref{cuba1}) predicts exponential (with respect to the
number of collisions) relaxation towards the equilibrium value $r_{\rm
eq}$ of $r$. The approach to equilibrium is governed by the factor
$(1-A)^m$ meaning that when $|A|\ll 1$ the effective number of
interactions after which the equilibrium is established (which is
proportional to the relaxation time) equals to $-1/[\ln(1-A)]$.

\subsection{Minimization of the relaxation time}

\label{relax}

Since we are interested in possibly shorter relaxation time, we need
to maximize $A$ over the unitary ${\cal V}$ [under condition
(\ref{brams})]. To do that, we first write $P_0\otimes\Pi_{E+\eps}$ in
a conveniently chosen matrix form in the energy eigenbasis

\BEA \label{long}
&&\hspace{-0.4cm}P_0\otimes\Pi_{E+\eps}=\\ \hspace{-0.4cm}&&\diag(\underbrace{...,1,...,1,...}_{n_{E+\epsilon}\,\, {\rm elements}}|\underbrace{...,0,...,0,...|\hspace{-0.5mm}\cdots\hspace{-0.5mm}|...,0,...,0,...}_{n\,\, {\rm sections}}),~~~~~~~
\label{mrr}
\EEA
where sections correspond to eigenvectors of $H$, and elements in
sections run over the ones of $H_B$.  Values are shown only for the
subspace given by $P_0\otimes\Pi_{E+\eps}+P_\eps\otimes\Pi_E$.  The
first section (denoted as $|...|$) there are $n_{E+\eps}$ unities. Then
come $n$ identical sections, each one contains $n_E$ zeroes.

To maximize $A$ we need to minimize $\langle 0|\,(\, {\rm tr}_{\rm
  B}{\cal V}\P_0\otimes \Pi_{E+\eps} {\cal V}^\dagger\,)|0\rangle$
over all possible ${\cal V}$s living in
$P_0\otimes\Pi_{E+\eps}+P_\eps\otimes\Pi_E$.  One can show that the
optimal unitary amounts to a permutation of the eigenvalues
(\ref{mrr}) (this fact can be shown similarly to the derivation
presented in (9--12) of the main text).  Now we note that the trace
over $\tr_B$ amounts to summing up elements in each section. So the
element $\langle 0|\,(\, {\rm tr}_{\rm B}{\cal V}\P_0\otimes
\Pi_{E+\eps} {\cal V}^\dagger\,)|0\rangle$ will be the sum of the
elements of the first section in the permuted diagonal in
(\ref{long}). The optimal permutation will thus be the one which takes
out of the first section as much unities as possible. Therefore, if
$n_{E+\eps}$ (the number of unities) is $<$ than $n_E n$|the number of
zeroes|then it is possible to move all unities, making $\min_{{\cal
    V}}\left\{\langle 0|\,(\, {\rm tr}_{\rm B}{\cal V}\P_0\otimes
  \Pi_{E+\eps} {\cal V}^\dagger\,)|0\rangle\right\}=0$.  Otherwise,
the latter quantity will be $n_{E+\eps}-n_E n$, leading us to the
following formula: \BEA
\label{blumkin}
A_{\rm max}= \frac{r_E\, {\rm min}\left[\, n_{E+\eps}, n_{E}n\,\right]}{r_{\rm eq}n}.
\EEA

\subsection{One-shot relaxation}
\label{shot}

Now the shortest relaxation corresponds to just one collision
and it is reached for $A=1$, e.g., $r_E^{[1]}=r_{\rm eq}$ and
$n^{[1]}_{E+\eps}=n-1$ in (\ref{blumkin}). Then the corresponding
unitary operator ${\cal V}$ is the SWAP operation. The relaxation time
in this case amounts to one inter-collision time.

However, in this case the bath should consist of particles that have
the same energy gap as the system. This is not a realistic model for
the bath. Below we study a fully realistic bath model and show that
although the relaxation time in that situation is larger than a single
collision time, it still allows to conclude that the Carnot efficiency
can be reached at a large power.

\subsection{Relaxation time for realistic bath}

Let us work out (\ref{blumkin}) for a realistic example of the bath. We
assume that the bath particle amounts to $L\gg 1$ independent two-level
systems. Each such system has energies $0$ and $\zeta$. Thus the bath
particle has energies $0,\zeta,2\zeta,...,L\zeta$. Each bath energy level $E$
is degenerate
\BEA
n_E=\frac{L!}{(E/\zeta)!(L-E/\zeta)!}
\EEA
times. Provided that
\BEA
\label{or}
n_{E+\eps}> n_{E}n,
\EEA
we obtain
\BEA
\label{toro}
A_{\rm max}=(1+n e^{-\beta\epsilon})\, \frac{n_E e^{-\beta E}}{Z}.
\EEA
Since we want a larger $A_{\rm max}$, we take
\BEA
\label{coca}
E=L/(e^{\beta\zeta}+1).
\EEA
Hence using the Stirling's formula
$L!\simeq \sqrt{2\pi L}\, L^L e^{-L}$ and noting that $Z=(1+e^{-\beta\zeta})^L$ we get from (\ref{toro})
\BEA
A_{\rm max}
\simeq (1+n e^{-\beta\epsilon})\,\sqrt{\frac{1+e^{-\beta\zeta}}{L}}.
\label{butul}
\EEA
We work out (\ref{or}) via the Stirling's formula and obtain from (\ref{or}, \ref{coca})
\BEA
\label{cobra}
\frac{\ln[n]}{L}+h_2[\frac{1}{e^{\beta\zeta}+1}]< h_2[\frac{1}{e^{\beta\zeta}+1}+\frac{\epsilon}{L\zeta}],
\EEA
where $h_2[x]=-x\ln x-(1-x)\ln (1-x)$.

Let us specify $\epsilon$ as [$\mu$ is a parameter]
\BEA
\label{ku}
\epsilon= \mu T\ln[n], ~~ \mu>1.
\EEA
Putting this into (\ref{cobra}) we get
\BEA
\label{du}
\frac{\ln[n]}{L}+h_2[\frac{1}{e^{\beta\zeta}+1}]< h_2[\frac{1}{e^{\beta\zeta}+1}+\frac{\mu}{\beta\zeta}\,\frac{\ln[n]}{L}].
\EEA
Provided that $\mu>1$, (\ref{du}) can be satisfied for sufficiently
small (but finite) $\frac{\ln[n]}{L}$ and sufficiently large $\zeta>0$.
Note that (\ref{du}) never holds for $\mu\leq 1$.

Returning to (\ref{butul}) we see from (\ref{ku}) that for $\ln[n]={\cal O}(L)\gg 1$
\BEA
\label{gru}
A_{\rm max}={\cal O}(\frac{1}{\sqrt{\ln[n]}}),
\EEA
which means that the relaxation time scales as ${\cal O}(\sqrt{\ln[n]})$.

Consider now the opposite [to (\ref{or})] case
\BEA
\label{oor}
n_{E+\eps}< n_{E}n,
\EEA
where
\BEA
\label{otoro}
A_{\rm max}=(1+\frac{1}{n e^{-\beta\epsilon}})\, \frac{n_{E+\epsilon} e^{-\beta (E+\epsilon)}}{Z}.
\EEA
Choosing
\BEA
\label{ococa}
E+\epsilon=L/(e^{\beta\zeta}+1),
\EEA
we get [cf. (\ref{butul})]
\BEA
A_{\rm max}
\simeq (1+\frac{1}{n e^{-\beta\epsilon}})\,\sqrt{\frac{1+e^{-\beta\zeta}}{L}}.
\label{obutul}
\EEA
If we specify [cf. (\ref{ku})]
\BEA
\label{oku}
\epsilon= \nu T\ln[n], ~~ \nu<1,
\EEA
then (\ref{oor}) reads [cf. (\ref{du})]
\BEA
\label{odu}
h_2[\frac{1}{e^{\beta\zeta}+1}]< \frac{\ln[n]}{L}
+
h_2[\frac{1}{e^{\beta\zeta}+1}-\frac{\nu}{\beta\zeta}\,\frac{\ln[n]}{L}].
\EEA
This relation holds for $\nu<1$, sufficiently
small (but finite) $\frac{\ln[n]}{L}$ and sufficiently large $\zeta>0$.

Hence from (\ref{odu}, \ref{otoro}) we return to the same conclusion
(\ref{gru}).

\subsection{Relations with the main text}

In the main text|see in particular (16, 19)| we studied the relaxation
of the optimal engine E that has the energy spectrum described at the
beginning of section IV. More specifically, in the main text we needed
two different relaxation scenario: E with energy gap
$\epsilon^{[1]}\simeq T_1\ln[n]$ relaxes on a thermal bath at
temperature $T_2$ (where $T_1>T_2$), and conversely E with energy gap
$\epsilon^{[2]}\simeq T_2\ln[n]$ relaxes on a thermal bath at
temperature $T_1$; see (18) of the main text in this context.

Now the first case corresponds to (\ref{ku}, \ref{du}, \ref{gru}), while
the second case to (\ref{oku}, \ref{odu}, \ref{gru}). In both cases we
get that for a realistic thermal bath (but with tuned system-bath
interactions) the relaxation time amounts to ${\cal O}(\sqrt{\ln[n]})$;
see (\ref{gru}).

\section{Entanglement generation during relaxation}
\label{enta}

It is important to understand to which extent quantum are the
polynomial-time relaxation mechanisms discussed above. First of all,
how much system-bath entanglement has to be generated in the course
of relaxation? Recall that entanglement is an essentially quantum
resource and a sizable amount of entanglement would pose an additional
limitation on approaching the Carnot limit. Such additional
limitations are likely to be absent, as we demonstrate below.

Note that though the collisional relaxation scheme do not assume any
entanglement between the system and bath particle both before and
after the collision, it can still imply that some amount of
entanglement is generated during the collision. Nevertheless, in $\ln
n\to\infty$ limit (where the machine achieves the Carnot limit for the
efficiency) the amount of entanglement goes to zero linearly with $\ln
n$; recovering, thus, the classical nature of our setup in
thermodynamic limit.

The fact that the system and bath get entangled during the relaxation
(especially in the mesoscopic regime) implies that the dynamics of the
system is non-Markovian \cite{shres}.

Since the simplest case, the one-shot relaxation
(see section (\ref{shot})), has all the traits of the
phenomenon, we will show the above assertion on that particular example.

Say in the first relaxation step the system E starts with the diagonal
state $\rho=\frac{1}{1+n u_1}\diag(1,u_1,...,u_1)$ and with
hamiltonian $H_2=\diag(0,\eps^{[2]},...,\eps^{[2]})$. Where $u_1$ and
$u_2$ (and, through it, $\eps^{[2]}$) are determined from Eq. (16) of
the main text.  The whole relaxation process is but a SWAP operation
between E and one bath particle B which has the same hamiltonian $H_2$
and is in a thermal state with bath temperature $T$:
$\sigma_B=\frac{1}{1+n u_2}\diag(1,u_2,...,u_2)$.

The relaxation progresses autonomously -- no energy flow in or
out happens (otherwise one would need a third system to accept/give
energy and a control to switch on and off the interactions, while
the relaxation is supposed to be a probably prearranged but a
spontaneous process). To that end, the interaction between E and B,
$H_{EB}$, must satisfy \cite{mityug}
\BEA
\label{apo}
[H_{EB},\,H_E\otimes 1_B+1_E\otimes H_B]=0.
\EEA
Hence for $H_{EB}$ to be nontrivial it must act within the direct sum
$\cal{H}_D$ of two nonintersecting degenerate subspaces of $H_E\otimes 1_B+1_E\otimes H_B$
(the one spanned on eigenvectors with energy $\eps^{[2]}$ and the other
-- on eigenvectors with energy $2\eps^{[2]}$). The unitary preforming
SWAP also lives in that subspace, and, thus, can be generated by a
suitably chosen $H_{EB}$. So the relaxation is executed by a one-parametric
continuous family of unitaries $U(t)$ that live in $\cal{H}_D$ and satisfy
\BEA \label{unitar}
U(0)=1,\quad U(t_{rel})={\rm SWAP}
\EEA
where $t_{rel}$ is the duration of the collision=relaxation time.

Now, we introduce the following quantity
\BEA \label{ezattu}
E=\min_{U(t)}\left\{ \max_{t}\left[ \mathcal{N}\left( U(t)\rho\otimes\sigma_B U^{\dagger}(t) \right) \right] \right\};
\EEA
where $\mathcal{N}$ is the entanglement negativity \cite{vidal,bacatr}, which
measures the entanglement between the systems.

\begin{figure}[ht]
\includegraphics[width=8cm]{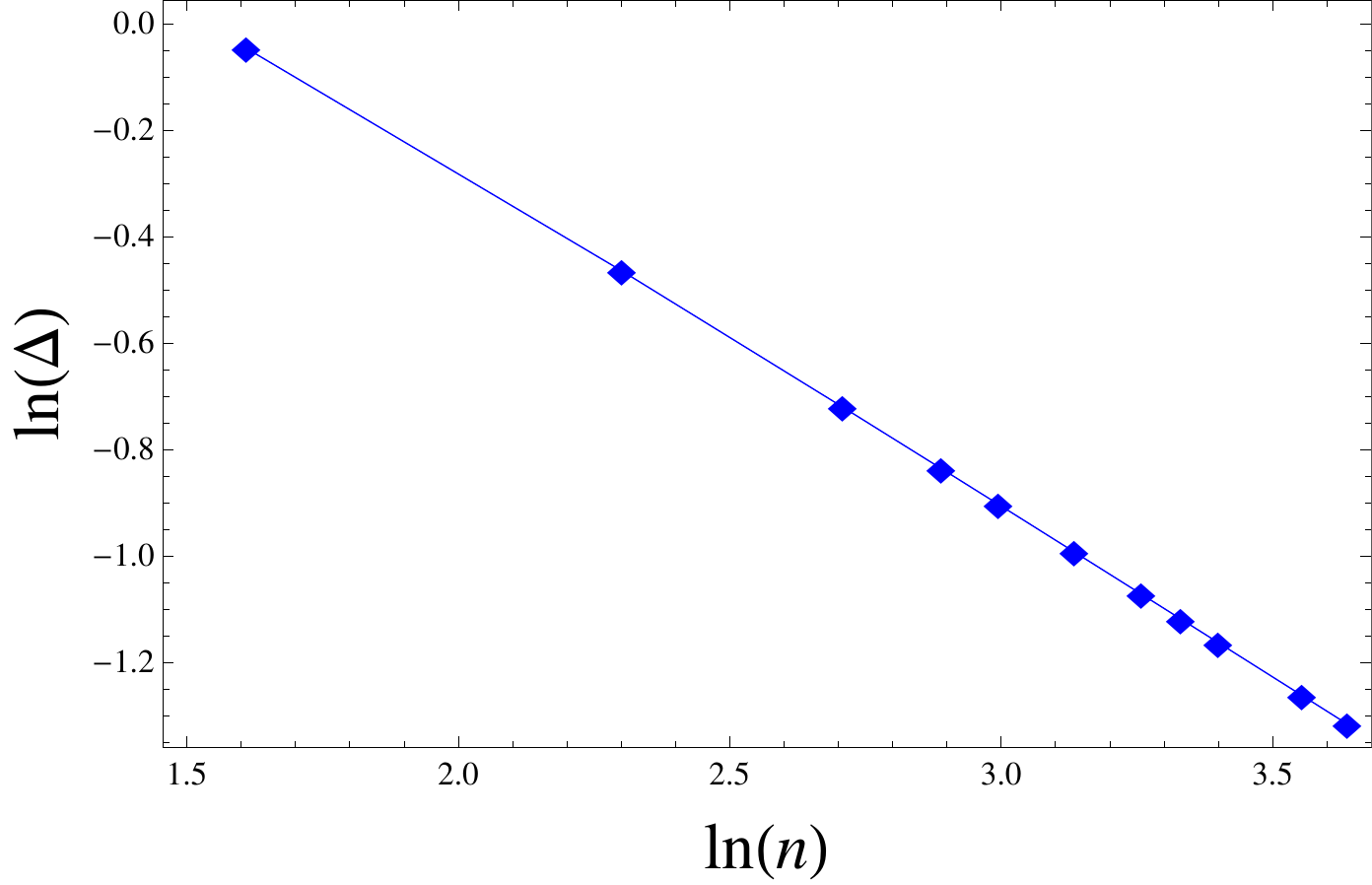}
\caption{The dependence of $\ln\left(\Delta(n,T_1)\right)$ on
  $\ln\left(n\right)$ for $T_1=2$. The dependence is linear within the
  numerical errors. The slope of the line $b\approx 0.65$; hence
  $\Delta(n,T_1)\sim n^{-b}$.}
\label{nkar}
\end{figure}

The quantity $E$ in (\ref{ezattu}) indicates the very necessity of
entanglement generation, since it finds the maximum over a process and
then takes the minimum of the maxima over all processes consistent
with constraints (\ref{unitar}). So if it is zero, then one can find a
process which proceeds without entangling the system and the bath
particle, while if it is positive, any process will reach a point
during its runtime when it starts to entangle the parties involved.
Note that when calculating $E$ we do not take into direct account the
constraint generated by (\ref{apo}).

As applied to our problem, we determine the quantities
$u_{1,2}(n,T_1,T_2)$ from Eq. (16) of the main text, then plug the
resulting states $\rho$ and $\sigma_B$ in (\ref{ezattu}) and perform
the min-max optimization. The resulting quantity is the indicator
(\ref{ezattu}) as a function of $n$, $T_1$, and $T_2$ (denote it by
$E_o=E_o(n,T_1,T_2)$).

\begin{figure}[ht]
\includegraphics[width=8cm]{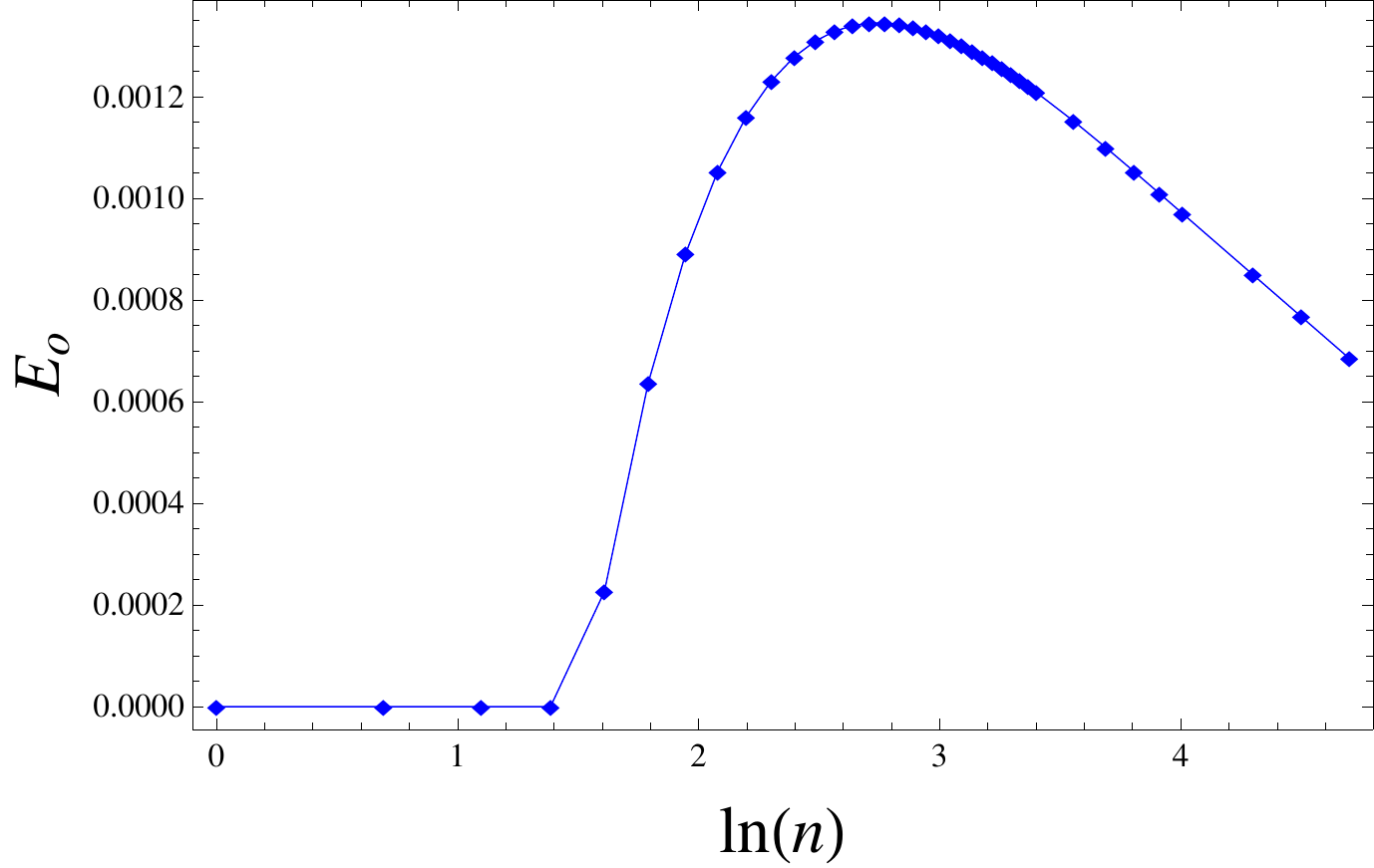}
  \caption{The dependence of $E_o(n,T_1,T_2)$ on
    $\ln\left(n\right)$ for $T_1=2$, $T_2=1$. As is clearly seen, the
    dependence becomes linear for $n\gtrsim30$.}
\label{nkarik}
\end{figure}

There is no entanglement when $T_2=T_1$ since the states $\rho$ and
$\sigma_B$ become identical and the SWAP does not alter them so that
the joint state remains constant and factorized. There the minimal
eigenvalue of the partially transposed density matrix is strictly
positive, so the continuity implies that there is a whole interval
$(T_1-\Delta(n,T_1),T_1)$ of $T_2$ that $E_o(n,T_1,T_2)=0$. This is
indeed certified by the numerics which also enables to calculate the
interval length $\Delta$ as a function of $n$ and $T_1$. It turns out,
that for any fixed $T_1$, $\Delta$ decreases with $n$. Also, the
numerical data (see Fig.~(\ref{nkar})) suggest that $\Delta\to0$ as
$n\to\infty$.

If we now fix the temperatures of the baths and calculate the
$E_o(n,T_1,T_2)$ as a function of $n$, we will indeed see that for small
$n$'s there is no negativity. Then, starting from some $n$, it starts
to appear and grow. But unexpectedly the negativity passes through a
maximum and starts to decrease linearly with $\ln n$ (see Fig.~(\ref{nkarik})).
So, as $\ln n\to\infty$, the necessity to generate entanglement drops out.
The latter is the limit where the engine's efficiency
approaches the Carnot value. So in the asymptotic limit we have both
$\eta\to\eta_C$ and $E_o\to0$ for any $T_2<T_1$; where the latter means
that although entanglement is necessarily generated during the relaxation for
almost any unequal temperatures of baths, its amount goes to zero so that
the dynamics becomes essentially classical.


\begin{thebibliography}{99}
\bibitem{foxnote} Since the full Hamiltonian is not uknown this
  measurement will at least take $\hbar/(\delta\epsilon)$ seconds due
  to the energy-time uncertainty principle, where $\delta\epsilon$ is
  the precision by which the ground-state energy is measured; see
  \cite{ahar}. This time is however a small overhead to the search
  time.
\end{thebibliography}
\begin{thebibliography}{99}
\bibitem{bacatr} Negativity is the sum of negative eigenvalues of the
  joint density matrix after a partial transposition. When it is
  positive then one knows for sure that there is some entanglement
  present, while one cannot guarantee that there is no entanglement
  when it is zero.
\end{thebibliography}
\end{document}